\documentclass{PoS}

\title{Dileptons, Charm and Charmonium at Finite Temperature and Chemical
Potential}

\ShortTitle{Dileptons and Charm(onium) at Finite $\mu_q$}

\author{\speaker{Ralf Rapp}
\\
Cyclotron Institute and Physics Department, Texas A\&M University,
College Station, TX 77843-3366, USA\\
E-mail: \email{rapp@comp.tamu.edu}}

\abstract{We discuss how dileptons, open charm and charmonia may be
utilized in heavy-ion collisions to extract information specific to 
hot and dense matter at finite quark chemical potential, $\mu_q$. For 
each observable we briefly discuss underlying theoretical frameworks 
and the current status in interpreting available heavy-ion data at 
SPS and RHIC energies. Low-mass dileptons are particularly sensitive
to baryonic medium effects in spectral modifications of the $\rho$
meson, and may serve as an accurate measure of the fireball lifetime. 
In the open-charm sector, observable signals may be generated by a 
``critical'' enhancement of scattering rates via $t$-channel exchange 
of a soft $\sigma$ mode. For charmonia, finite-$T$ potential models 
could be used to extrapolate color-screening effects to finite $\mu_q$ 
to facilitate a quantitative evaluation of dissociation rates in the 
medium.}

\FullConference{5th International Workshop on Critical Point and Onset of Deconfinement - CPOD 2009,\\
		 June 08 - 12 2009\\
		 Brookhaven National Laboratory, Long Island, New York, USA}

\begin{document}

%%%%%%%%%%%%%%%%%%%%%%%%%%
\section{Introduction}
\label{sec_intro}
\begin{figure}[!b]
\begin{center}
\includegraphics[width=0.60\textwidth,angle=-90]{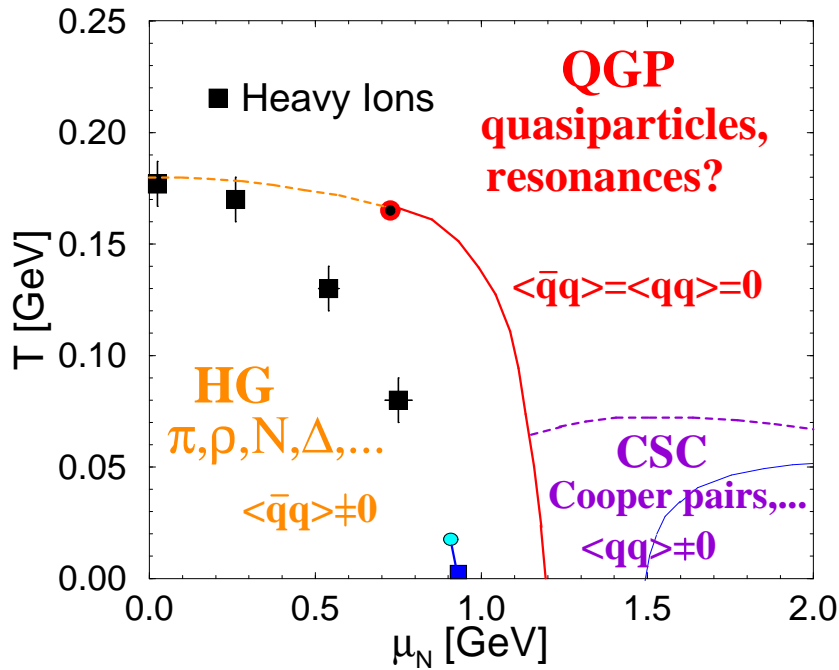}
\end{center}
\caption[]{Schematic diagram of the QCD phase structure in the $\mu_q$-$T$
plane, as characterized by the lowest dimension quark condensates, i.e.,
the chiral quark-antiquark and the diquark one (Cooper pairs). The former
is believed to prevail in the low-$T$ and -$\mu_q$ hadronic world while
the latter occurs at the Fermi surface of cold dense quark matter. Key to
investigating the phase structure is the identification of the relevant
excitations in the respective phases, as indicated in the figure. The 
``data'' points are extracted from hadro-chemical analysis of the final
state observed in heavy-ion collisions~\cite{BraunMunzinger:2003zd,
Becattini:2003wp}.
}
\label{fig_phasedia}
\end{figure}

%%%%%%%%%%%%%%%%%%%%%%%%%%
Rigorous information on the QCD phase diagram, schematically depicted
in Fig.~\ref{fig_phasedia}, is scarce. At finite temperature ($T$) and
vanishing quark chemical potential ($\mu_q=0$), lattice QCD (lQCD) 
computations predict the existence of a rapid cross-over transition 
from hadronic matter to a Quark-Gluon Plasma at a (pseudo-) critical 
temperature of $T_c=(180\pm20)$~MeV~\cite{Cheng:2007jq,Aoki:2006br}. 
Heavy-ion collisions at SPS and RHIC energies have shown that the produced
medium exhibits a high degree of thermalization and that the achieved
temperatures are on the order of $T_c$ and above. This provides a firm 
basis for studying QCD matter in the laboratory. At finite $\mu_q$ and
vanishing $T$, there are compelling arguments that the high-density limit
of QCD is in a color-superconducting phase (attractive quark-quark 
interaction plus Cooper theorem). Since this ground state, characterized 
by a non-vanishing diquark condensate ($\langle qq\rangle \ne 0$), is
qualitatively different from the chirally broken QCD vacuum 
($\langle \bar{q}q\rangle \ne 0$), it is natural to expect a first-order
transition at some intermediate chemical potential, $\mu_q^c\simeq400$~MeV.  
Finally, heavy atomic nuclei have been long used to quantify the 
properties and excitatins of the finite-density ground state at 
$\mu_q = (m_N-E_B)/3 \simeq 310$~MeV, indicated by the square box in 
Fig.~\ref{fig_phasedia}. 

Heavy-ion experiments performed over a wide range of collision energies 
can, in principle, cover a large region of the phase diagram, cf.~the
``data'' points in Fig.~\ref{fig_phasedia}. In particular, as indicated 
above, one ought to be able to determine phase changes, down to 
temperatures of about 100~MeV where the critical chemical potential may 
be around $\mu_q^c\simeq400$~MeV (it is, however, questionable 
whether heavy-ion experiments can reach into the color-superconducting 
phases, unless the zero-temperature quark pairing gap is well above 
100~MeV~\cite{Rapp:1999qa}; even if so, equilibration appears to be 
unlikely. More likely could be the production of the so-called 
quarkyonic phase~\cite{McLerran:2007qj} which may extend to higher $T$).
Of special interest is the occurrence of a critical second-order 
endpoint, whose existence is suggested by the cross-over transition 
found  in finite-$T$ lQCD and the putative 1.~order transition at $T=0$.
Suitable observables to study the QCD phase structure in heavy-ion 
collisions (HICs) may be roughly divided into two categories:
(1) bulk observables driven by the equation of state (EoS), including 
collective flow patterns of the most abundant particles ($\pi$, $K$, $p$) 
or fluctuation observables in $p_T$ spectra and hadrochemistry; 
(2) ``microscopic'' probes associated with specific quantum-number
channels, e.g., the vector current coupling to photons and dileptons. 
In some instances the physical origin of type (1) and (2) observables is 
closely related. E.g., electric-charge fluctuations are governed by the
electromagnetic (EM) susceptibility, $\chi_{\rm em}$, which can be 
expressed as the screening limit of the static EM correlation 
function, $\langle Q^2 \rangle - \langle Q\rangle^2 
= \chi_{\rm em} = \Pi_{\rm em}(q_0=0,q\to 0)$, while photon and
dilepton spectra are directly proportional to the imaginary part
of $\Pi_{\rm em}$, as discussed below.  
    
In this paper we focus on microscopic probes as realized via dileptons,
charm and charmonia. Corresponding observables are often associated with
``hard probes'', due to a large momentum transfer associated with their 
initial production (e.g., $|q^2| \ge 4m_c^2 \simeq6$~GeV$^2$). We will 
argue, however, that all of the above 3 probes can provide valuable 
information on relatively ``soft'' modes in the medium, at the 
temperature scale or below. For dileptons (Sec.~\ref{sec_dilep}), this 
relates to the in-medium $\rho$-meson spectral function as reflected in 
thermal radiation at low invariant masses ($M\sim {\cal O}(T)\le m_\rho$). 
In the open-charm sector (Sec.~\ref{sec_charm}), one can study mechanisms 
of thermalization or, more generally, transport properties within a 
diffusion equation via modifications of charmed hadron $p_T$ spectra and 
elliptic flow (governed by elastic scattering at typical momentum 
transfers $|q|\sim {\cal O}(gT)$). Finally, for charmonia 
(Sec.~\ref{sec_charmonium}), the key to understanding their in-medium 
bound-state properties lies in color-screening effects (at the Debye 
scale $m_D\sim{\cal O}(gT)$) as well as inelastic dissociation 
reactions (at the binding-energy scale). In each sector, based on current 
insights, we will try to identify promising directions of future 
investigations specific to the situation of finite $\mu_q$ and/or the 
putative critical point.         
Concluding remarks are collected in Sec.~\ref{sec_concl}.

%%%%%%%%%%%%%%%%%%%%%%%%%%%%%%%%%
\section{Low-Mass Dileptons: $\rho$-Meson Spectroscopy}
\label{sec_dilep}
%%%%%%%%%%%%%%%%%%%%%%%%%%%%%%%%
The basic quantity to calculate thermal emission spectra of
EM radiation from hot and dense matter is the retarded correlation 
function of the hadronic EM current, $j^\mu_{\rm em}$,
\begin{equation}
\Pi_{\rm em}^{\mu \nu}(M,q;\mu_B,T) = -i \int d^4x \ e^{iq\cdot x} \
\Theta(x_0) \ \langle[j_{\rm em}^\mu(x), j_{\rm em}^\nu(0)]\rangle_T \ .
\end{equation}
Its imaginary part (EM spectral function) directly figures into the 
differential production rates of dileptons ($l^+l^-$) and photons 
($\gamma$),
\begin{eqnarray}
\frac{dN_{ll}}{d^4xd^4q} &=& -\frac{\alpha_{\rm em}^2}{\pi^3} \ 
\frac{L(M)}{M^2} \ f^B(q_0;T) \  {\rm Im}~\Pi_{\rm em} (M,q;\mu_B,T) 
\label{Rll}
\\ 
q_0\frac{dN_{\gamma}}{d^4xd^3q} &=& -\frac{\alpha_{\rm em}}{\pi^2} \
       f^B(q_0;T) \  {\rm Im}~\Pi_{\rm em}(M=0,q;\mu_B,T) \ ,
\label{Rgam}
\end{eqnarray}
respectively ($f^B$ is the thermal Bose distribution and $L(M)$ a 
final-state lepton phase-space factor relevant close to the dilepton
threshold). In the vacuum, the low-mass regime ($M\le 1$~GeV) of 
$\Pi_{\rm em}$ is essentially saturated by the light vector mesons 
$\rho$, $\omega$ and $\phi$. Within the vector-dominance model (VDM) 
the EM spectral function is directly proportional to the vector-meson
spectral functions,  
\begin{equation}
{\rm Im}~\Pi_{\rm em} \sim [ {\rm Im}~D_\rho  +
\frac{1}{9} {\rm Im}~D_\omega + \frac{2}{9} {\rm Im}~D_\phi ] \ . 
\label{vdm}
\end{equation}
Thus, if VDM remains valid in the medium (see, e.g., 
Ref.~\cite{Harada:2003jx} for an alternative scheme), low-mass dilepton 
spectra mostly probe in-medium modifications of the $\rho$ meson, which 
have been studied rather extensively in the literature, see, e.g., 
Refs.~\cite{Rapp:2009yu,Leupold:2009kz} for recent reviews.  

It turns out that low-mass thermal EM radiation in HICs dominantly
emanates from the hadronic phases of the collisions, even at RHIC
energies~\cite{David:2006sr}. It is therefore in order to study 
hadronic medium effects on the $\rho$ propagator,
\begin{equation}
D_\rho(M,q;\mu_B,T) = 
[ M^2-{m_\rho^{(0)}}^2-\Sigma_{\rho\pi\pi}-\Sigma_{\rho B}-
  \Sigma_{\rho M} ]^{-1} \ , 
\end{equation}
encoded in selfenergy insertions, $\Sigma_\rho$, induced by interactions
with particles in the heat bath. These may be classified as 
(a) medium modifications of the pion cloud, $\Sigma_{\rho\pi\pi}$,
due to pion rescattering (most notably on baryons) and thermal Bose 
enhancement~\cite{Urban:1999im}; 
(b) direct $\rho$-baryon couplings~\cite{Friman:1997tc}, e.g., 
$\rho+N\to \Delta, N(1520), N(1720)$, etc.; 
(c) direct interactions of the $\rho$ with mesons, e.g., 
$\rho+\pi\to \omega,a_1,...$ or $\rho+K\to K_1,...$ etc.~\cite{Rapp:1999qu}.
The interactions are usually modeled by effective hadronic Lagrangians 
which satisfy basic constraints from EM gauge invariance and (mostly 
for pions) chiral symmetry. The free parameters (coupling constants and 
formfactor cutoffs to account for finite-size effects) can be 
constrained empirically by partial decay rates (e.g., 
$a_1\to \pi\rho, \pi\gamma$) or, more comprehensively, 
scattering data (e.g., $\pi N\to \rho N$ or photo-absorption cross 
sections).   
\begin{figure}[!t]
\begin{minipage}{0.5\linewidth}
\includegraphics[width=0.78\textwidth,angle=-90]{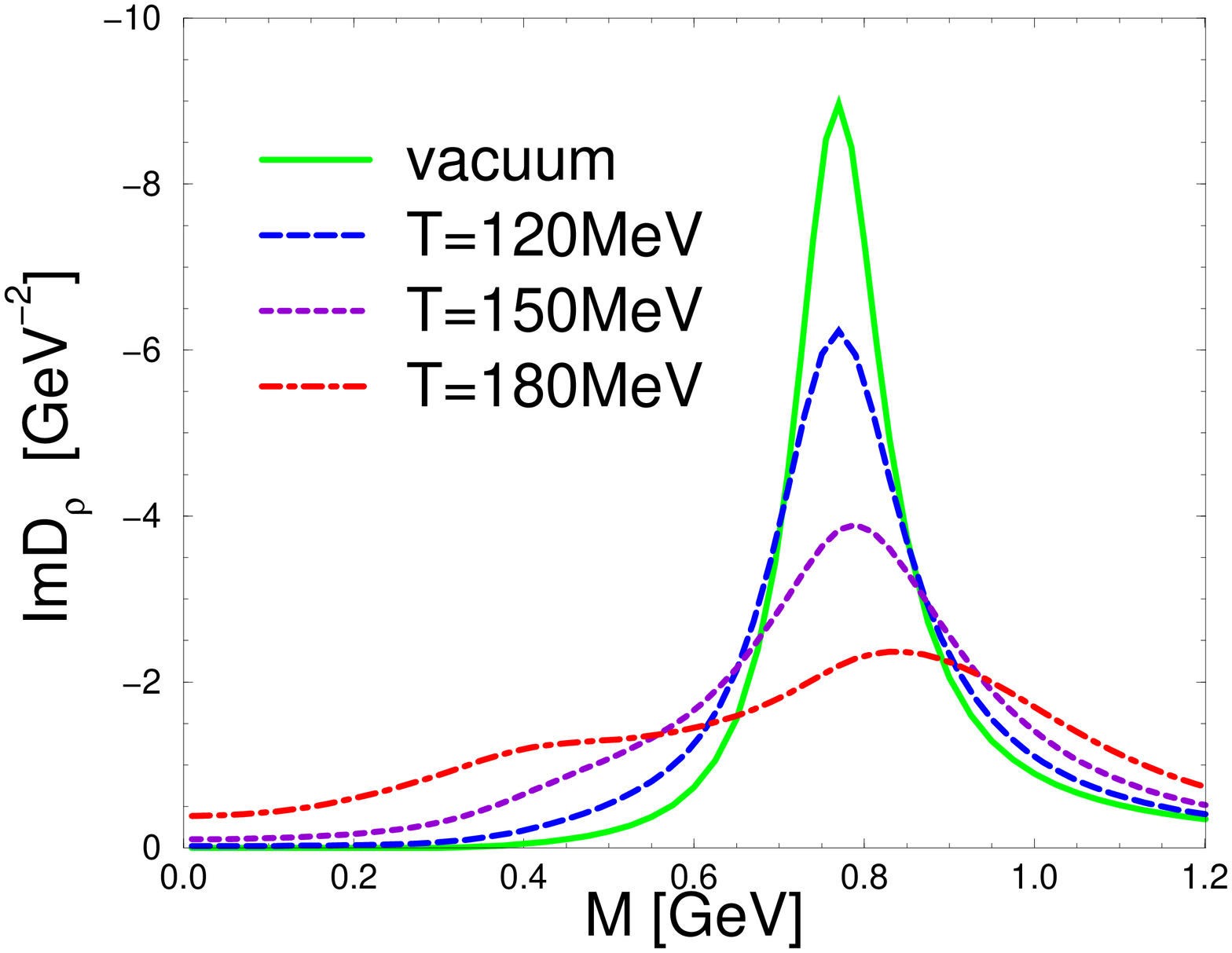}
\end{minipage}
\hspace{0.3cm}
\begin{minipage}{0.5\linewidth}
\vspace{-0.1cm}
\includegraphics[width=0.95\textwidth]{drdm2-Inx160.eps}
\end{minipage}
\vspace{0.1cm}
\caption[]{Left panel: $\rho$-meson spectral function in hot and dense 
hadronic matter at fixed $\mu_B=3\mu_q=330$~MeV (corresponding to
baryon densities $\varrho_B/\varrho_0=0.1,0.7,2.6$ at $T=120,150,180$~MeV,
respectively) and 3-momentum $q=0.3$~GeV~\cite{Rapp:1999us}. 
Right panel: 3-momentum 
integrated thermal dielectron rates in the isovector channel using the 
vacuum (dotted line) and full in-medium (solid line) $\rho$ spectral
function; the long-dashed, dash-dotted and short-dashed curves only 
include in-medium selfenergies due to either the in-medium pion cloud 
($\Sigma_{\rho\pi\pi}$) or direct $\rho$-baryon interactions 
($\Sigma_{\rho B}$) or direct $\rho$-meson interactions 
($\Sigma_{\rho M}$), respectively (the latter 2 include the free pion 
cloud as well).}
\label{fig_arho}
\end{figure}
The left panel of Fig.~\ref{fig_arho} shows an in-medium $\rho$ spectral
functions~\cite{Rapp:1999us} including all of the above components under
conditions roughly resembling HICs at the SPS. A strong broadening with
increasing matter density and temperature occurs, melting the resonance 
when extrapolated toward the phase boundary region.
The large low-mass enhancement becomes much more apparent in the
dilepton production rate, due to the Bose factor and photon propagator
in eq.~(\ref{Rll}), cf.~right panel of Fig.~\ref{fig_arho}. At 
$M\simeq0.4$~GeV, an order of magnitude enhancement over the rate 
from free $\pi\pi$ annihilation is predicted. Also note that the 
divergence of the rate for $M\to 0$ is required to produce a finite 
photon production rate, eq.~(\ref{Rgam}). Plotting the rate in terms 
of the 3 individual selfenergy contributions as introduced above, one 
clearly recognizes the prevalence of the baryon-driven medium effects. 
This may seem surprising since at SPS energies the observed 
pion-to-baryon ratio is about 5:1. However, in the interacting medium, 
most of the pions are ``stored'' in excited (meson and baryon) 
resonances; e.g., at ($\mu_B$,$T$)=(240,160)~MeV, the total baryon 
density, $\varrho_B\simeq0.8\varrho_0$ is quite comparable to the direct 
pion density, $\varrho_\pi\simeq0.9\varrho_0$ ($\varrho_0=0.16~$fm$^{-3}$). 

\begin{figure}[!b]
\begin{minipage}{0.5\linewidth}
\includegraphics[width=0.95\textwidth]{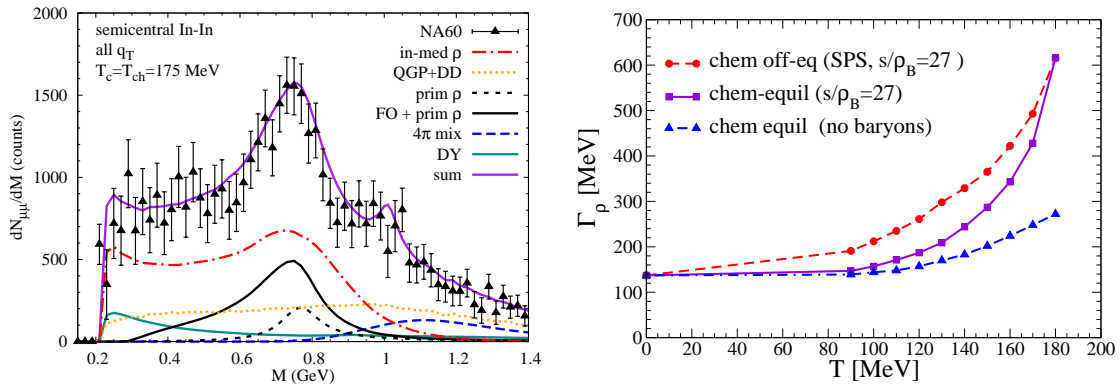}
\end{minipage}
\begin{minipage}{0.5\linewidth}
\includegraphics[width=0.95\textwidth]{Gam-emsf-SPS.eps}
\end{minipage}
\caption[]{Left panel: dimuon invariant-mass spectra (including experimental
acceptance) in semicentral In-In collisions at SPS energies ($E_{\rm lab}=
158$~AGeV); NA60 data~\cite{Arnaldi:2006jq} are compared to theoretical 
calculations based on the in-medium $\rho$ spectral function shown
in Fig.~\ref{fig_arho}. 
%The lower right panel illustrates the consequences of switching 
%off baryonic medium effects (``no bar") and using the vacuum $\rho$ 
%spectral function (``free").
Right panel: in-medium $\rho$ width as a function of temperature along
isentropic trajectories in the phase diagram starting from chemical
equilibrium at $(T,\mu_B)=(175,230)$~MeV (dots: fixed hadrochemistry
including chemical potentials for pions, kaons, etc.; squares: chemical 
equilibrium; triangles: without baryonic medium effects).}      
\label{fig_dndm}
\end{figure}

An application of the $\rho$ spectral function to recent NA60 low-mass 
dimuon data in In-In collisions at SPS~\cite{Arnaldi:2006jq} is shown 
in the left panel of Fig.~\ref{fig_dndm}. 
Upon convoluting the EM emission rates over an expanding fireball, the 
excess radiation is well described by the predicted in-medium $\rho$ 
line shape (QGP and primordial contributions are 
subleading)~\cite{vanHees:2007th}, see 
Refs.~\cite{Dusling:2007kh,Ruppert:2007cr,Bratkovskaya:2008bf} for
alternative calculations. This is also true for the excess dielectron 
data reported for central Pb-Au by CERES/NA45~\cite{Adamova:2006nu}. 
%The vacuum $\rho$, or meson-gas effects alone, clearly fail in describing
%the data (lower left). 
One can quantify the $\rho$ broadening by an approximate average width 
which amounts to 
$\bar{\Gamma}_\rho^{\rm med}\simeq$~(350-400)~MeV~$\simeq 
3~\Gamma_\rho^{\rm vac}$, realized at a representative temperature 
of $\bar{T}\simeq$~150-160~MeV, cf.~right panel of Fig.~\ref{fig_dndm}. 
This inevitably implies that, toward $T_c$, the $\rho$'s in-medium width 
becomes comparable to its mass, i.e., the resonance has indeed melted.
The absolute yield of the excess radiation is quite sensitive to the
total fireball lifetime, enabling a remarkably accurate measurement 
of the fireball lifetime for semicentral In-In collision, 
$\tau_{\rm FB}\simeq (6.5\pm1)$~fm/$c$. This tool could become
invaluable for detecting significant lifetime changes when approaching
the critical point and/or moving into the first-order regime with an
extended mixed phase (of course, it only works if the in-medium
spectral shape is under sufficient theoretical control). 

\begin{figure}[!t]
\begin{minipage}{0.5\linewidth}
\vspace{-1.2cm}

\includegraphics[width=0.92\textwidth]{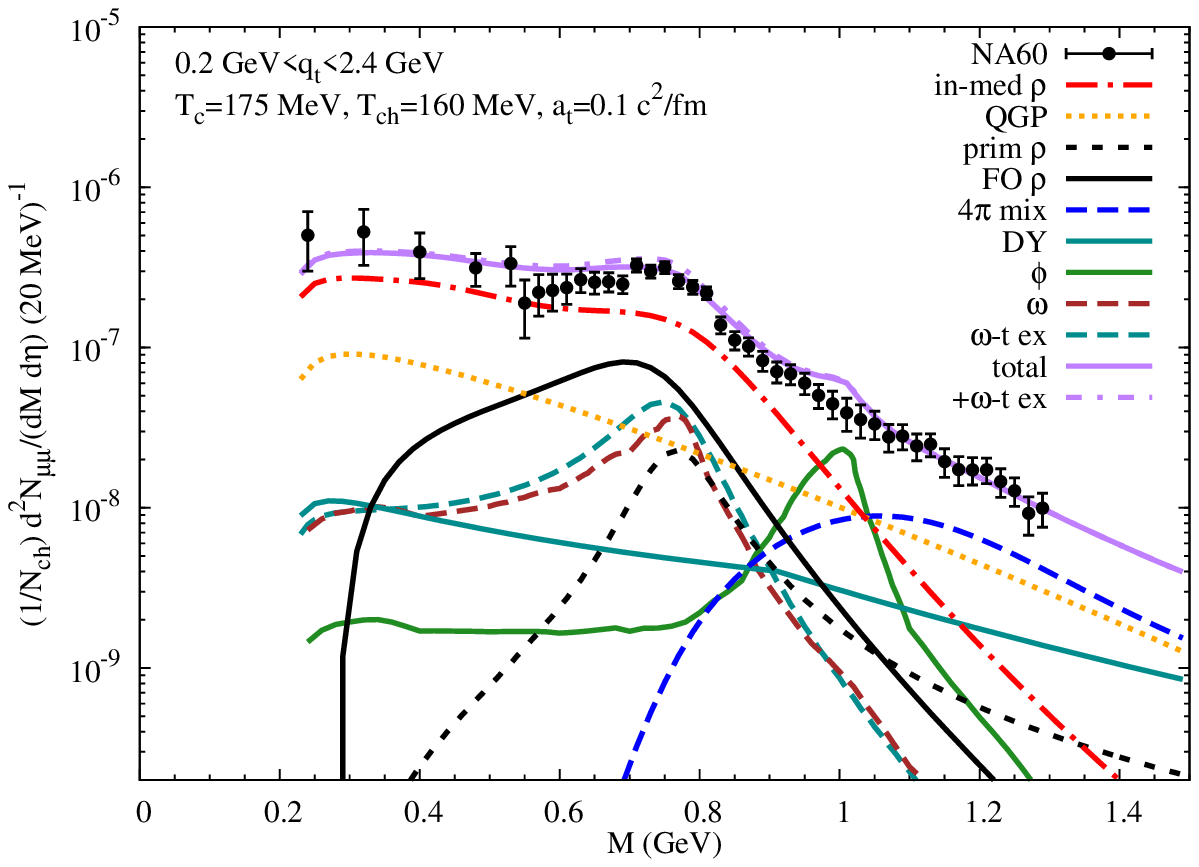}

\vspace{-0.2cm}
\includegraphics[width=0.92\textwidth]{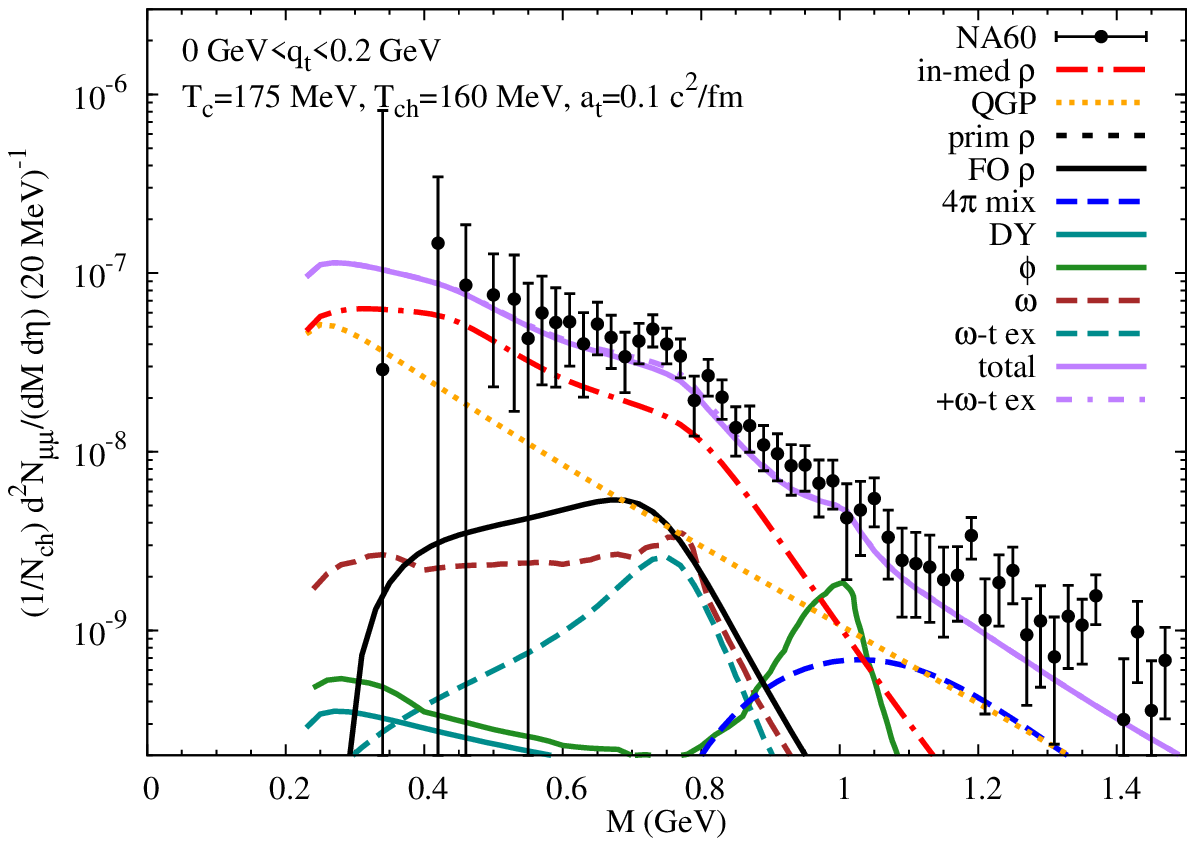}
\end{minipage}
\hspace{-0.4cm}
\begin{minipage}{0.5\linewidth}
\includegraphics[width=1.00\textwidth]{drdm-mu15-150-180.eps}
\end{minipage}
\caption[]{Left panel: acceptance-corrected dimuon invariant-mass spectra 
in semicentral In(158~AGeV)-In~\cite{Arnaldi:2008er,Arnaldi:2008fw} for
transverse pair momenta $q_t=0.2$-2.4~GeV (upper left) and 
$q_t=0$-0.2~GeV (lower left). 
Right panel: 3-momentum integrated dimuon thermal emission rates in
the isovector ($\rho$) channel at a baryon chemical potential
representative for SPS energies ($\mu_B=330$~MeV)~\cite{Rapp:1999us}.}
\label{fig_rate}
\end{figure}

The NA60 collaboration has recently taken another step forward by 
fully correcting their data for experimental 
acceptance~\cite{Arnaldi:2008er,Arnaldi:2008fw}. Upon integrating over
transverse momentum, the resulting invariant-mass spectra, 
$dN_{\mu\mu}/dMdy$, do justice to the notion of
Lorentz-{\em invariance}, i.e., transverse flow effects have been 
eliminated. Thus, one is essentially looking at the (average) emission 
rate from the medium, multiplied by the emitting 4-volume, cf.~left 
panels in Fig.~\ref{fig_rate}.    
This provokes a direct comparison to the theoretical input rates based 
on Ref.~\cite{Rapp:1999us}, augmented by the muon phase-space factor, 
$L(M)$, shown in the right panel of Fig.~\ref{fig_rate} for two 
temperatures. The resemblance of the in-medium hadronic rates and the 
NA60 spectra is rather astonishing, both in slope and shape. The former 
can, in principle, serve as a true thermometer, i.e. free from blue-shift 
contamination due to transverse flow. Essential to these
arguments is the prevalence of thermal radiation in the excess spectra
which is borne out of (i) the theoretical calculations, and
(ii) the complete lack of polarization in the measured angular 
distribution of the muon pairs~\cite{Arnaldi:2008gp}. The good overall 
agreement of theory and data furthermore corroborates that VDM 
stays intact even close to the phase boundary (the data also indicate 
that the $\phi$ does not radiate dileptons in the hadronic 
phase but decouples earlier~\cite{Adamova:2005jr,Arnaldi:2009wr}; this 
is, in fact, consistent with its relatively soft $p_T$ spectra).  

The importance of baryon-driven medium effects in the interpretation
of the SPS low-mass dilepton data naturally calls for studies at
lower collisions energies where even larger baryon compression 
may be achieved. 
\begin{figure}[!t]
\begin{minipage}{0.5\linewidth}
\includegraphics[width=0.8\textwidth,angle=-90]{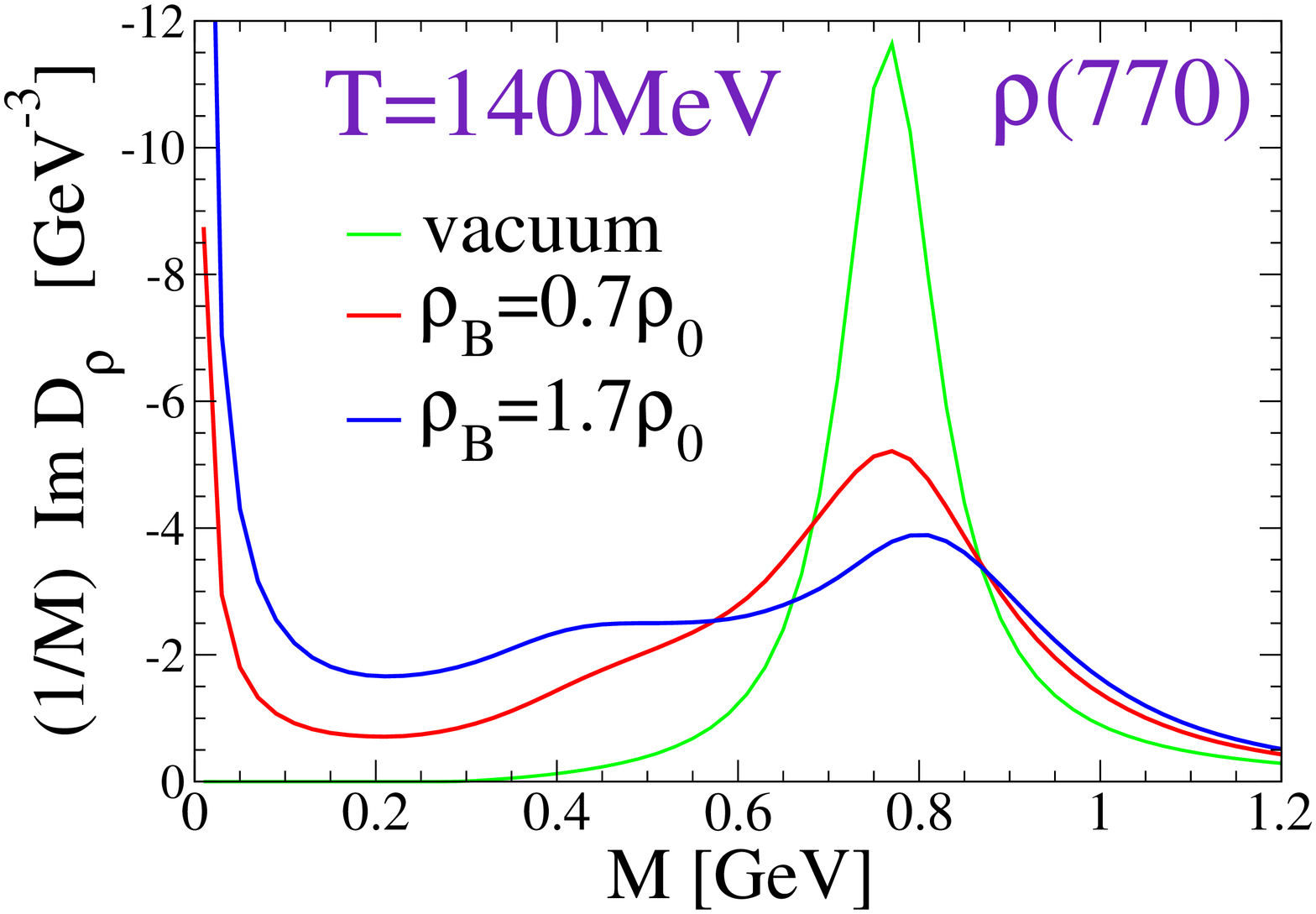}
\end{minipage}
\begin{minipage}{0.5\linewidth}
\includegraphics[width=0.8\textwidth,angle=-90]{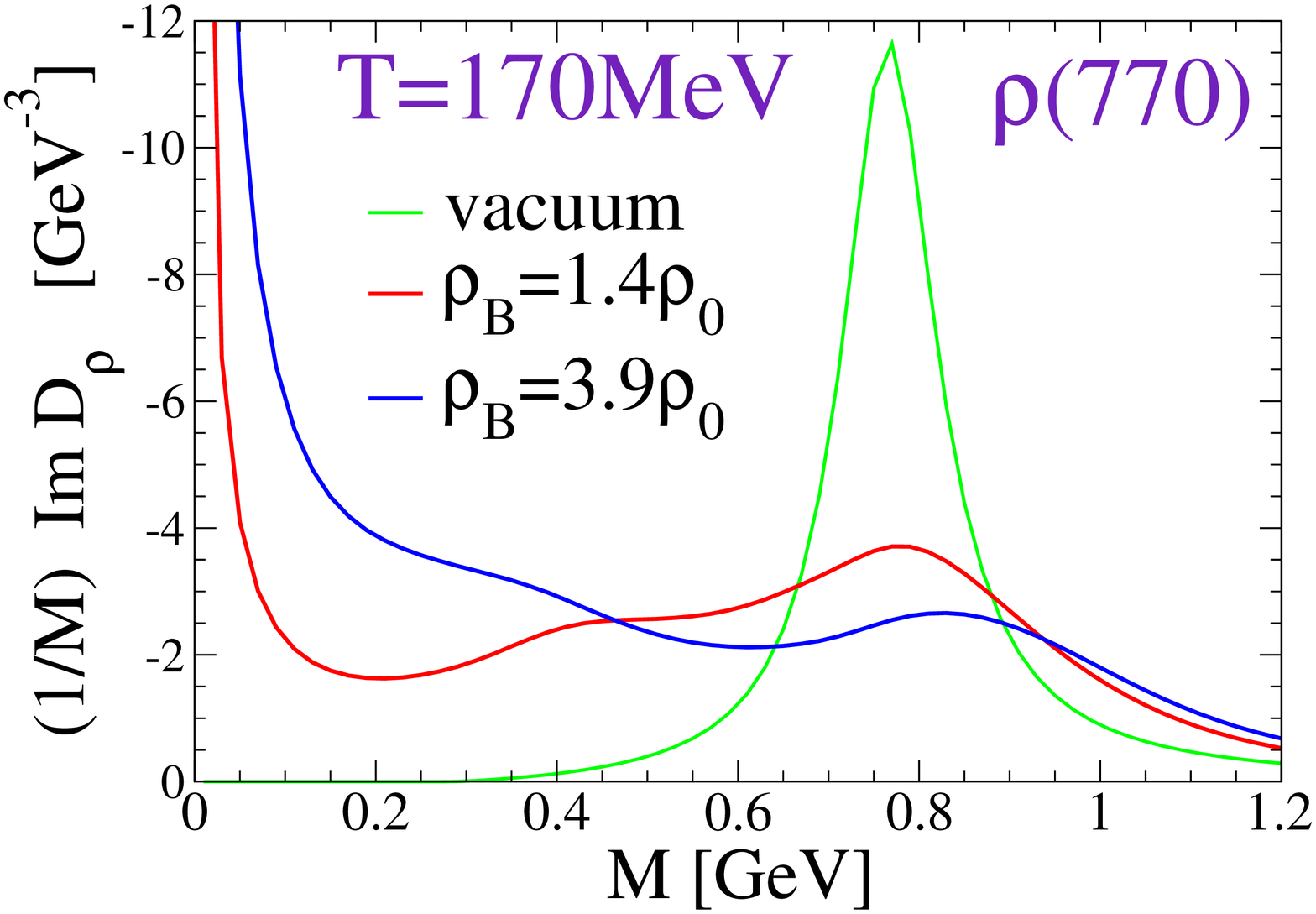}
\end{minipage}
\caption[]{$\rho$-meson spectral functions, weighted by a factor of 
inverse mass as figuring into the 3-momentum integrated dilepton rate, 
at two temperatures and for baryon densities representative for full SPS 
energy (red lines) and CBM energies (blue lines).}
\label{fig_arhom}
\end{figure}
Hadronic many-body calculations identify the mass regime around 
$M\simeq0.2$~GeV as the most sensitive one in the $\rho$ spectral 
function, with up to a factor of $\sim$2 enhancement under conditions
expected at the Compressed Baryonic Matter (CBM) experiment relative
to full SPS energy. A first glimpse at such an effect may have been
seen by CERES/NA45 in a 40~AGeV Pb-Au run~\cite{Adamova:2002kf}.     

\begin{figure}[!t]
\begin{minipage}{0.5\linewidth}
\includegraphics[width=1.0\textwidth]{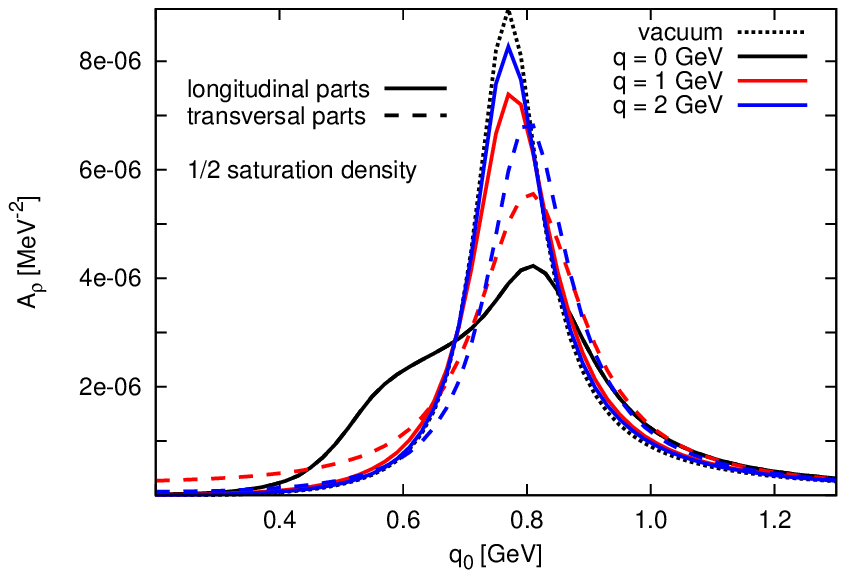}
\end{minipage}
\begin{minipage}{0.5\linewidth}
\vspace{-0.3cm}
\hspace{1.4cm}
\includegraphics[width=0.7\textwidth]{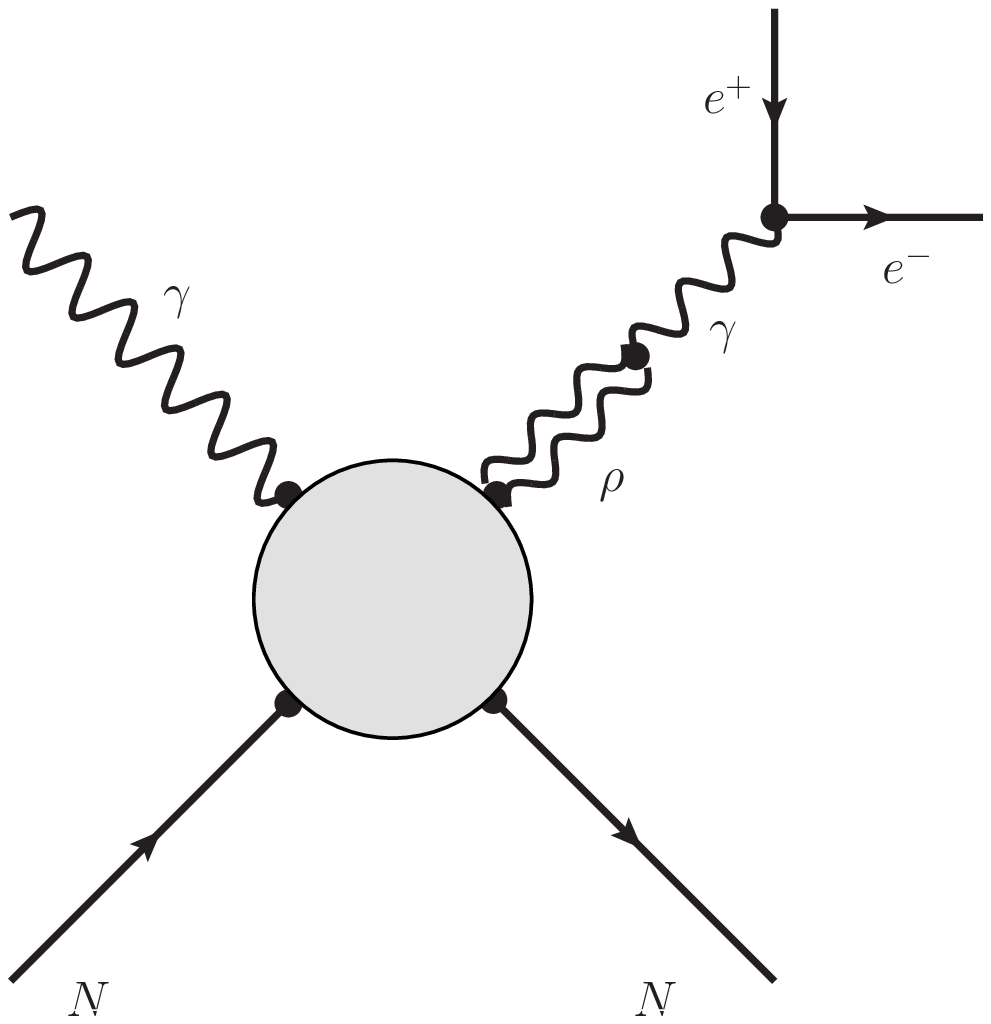}
\end{minipage}
\caption[]{Left panel: in-medium $\rho$ spectral function in cold 
nuclear matter at half saturation density for various 3-momenta (for
$q$>0, transverse and longitudinal modes split). 
Right panel: elementary amplitude for $\rho$ photo-production on the 
nucleon~\cite{Riek:2008ct}.
}
\label{fig_cold}
\end{figure}

A direct way to study baryon effects is provided by cold
nuclear matter, i.e., in atomic nuclei. The advantage over heavy-ion
experiments obviously lies in the essentially static matter environment,
which, however, is limited by nuclear saturation density ($\varrho_0$) 
and also exhibits significant spatial gradients. Nevertheless, the 
predicted medium effects on the $\rho$ spectral function are appreciable 
even at half saturation density, at least at small 3-momentum relative 
to the nuclear rest frame, see left panel of 
Fig.~\ref{fig_cold}~\cite{Rapp:1999us,Riek:2008ct}.
As in HICs, the dilepton final state is the cleanest way to probe medium
effects. The initial state in nuclear production experiments is, however, 
rather different: the $\rho$ has to be created by an external 
excitation (cf.~right panel of Fig.~\ref{fig_cold}), as compared to an 
approximately thermal medium in HICs. 
Thus, a good knowledge of the production process is mandatory, which
can be tested with proton targets. Two additional complications arise:
(a) an in-medium broadening leads to a reduction in the dilepton to
hadronic branching ratio, thus reducing the signal (in HICs the $\rho$
is ``continuously" regenerated in the interacting medium); 
(b) to provide the mass of the $\rho$ a rather energetic incoming
particle is needed which usually implies that the $\rho$ carries
significant 3-momentum; this enhances surface and/or escape effects 
thus reducing the in-medium signal as well.  
The latter point is presumably best dealt with using a photon beam
where all the incoming energy can be converted into mass.
Photo-production of dileptons has recently been studied by the CLAS 
collaboration at Jefferson Lab (JLab) using a variety of nuclear 
targets~\cite{Wood:2008ee}, with an incoming photon spectrum ranging
over $E_\gamma\simeq$~(1-3.5)~GeV. 
\begin{figure}[!t]
\begin{minipage}{0.5\linewidth}
\includegraphics[width=0.93\textwidth]{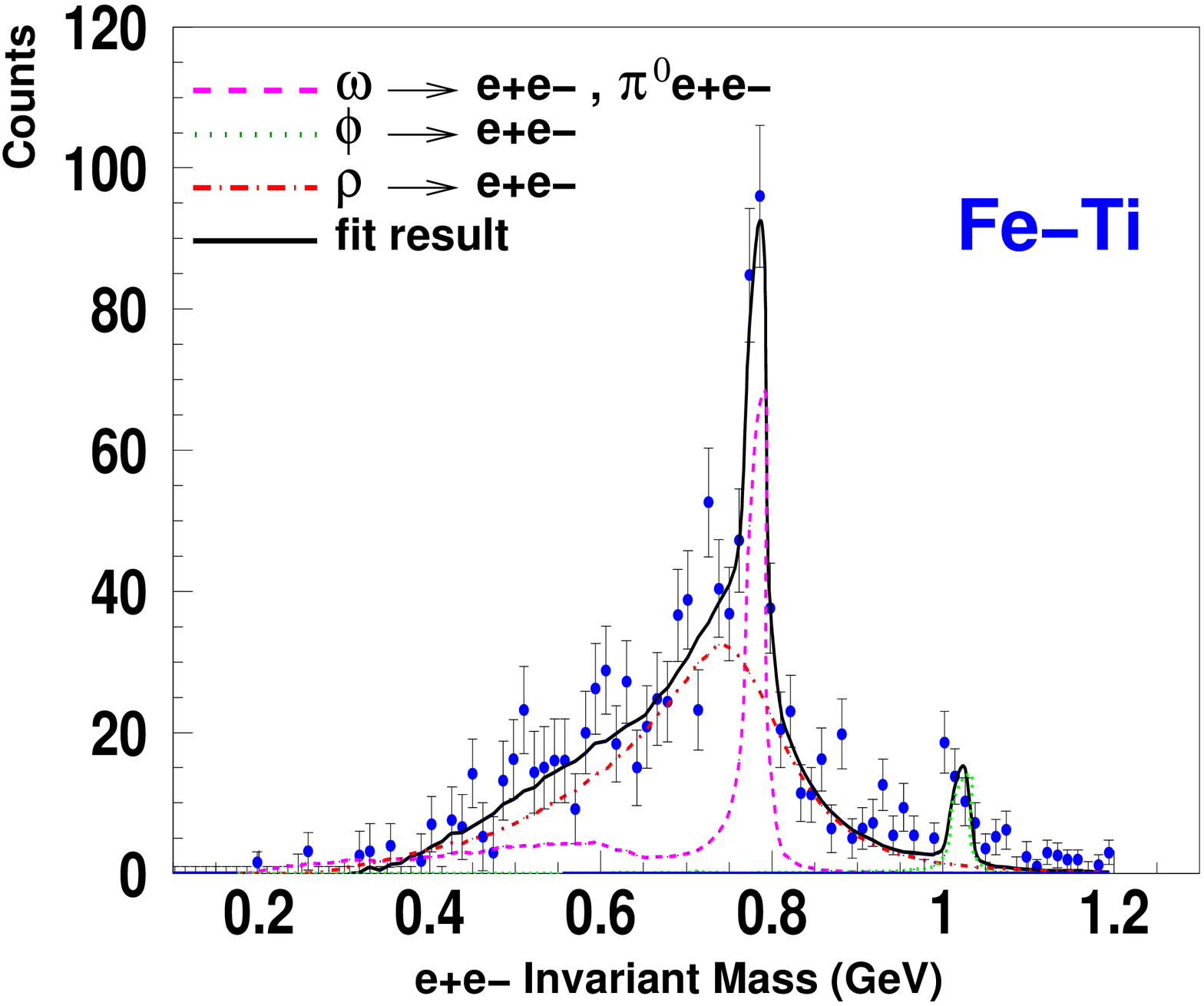}
\end{minipage}
\begin{minipage}{0.5\linewidth}
\includegraphics[width=1.0\textwidth]{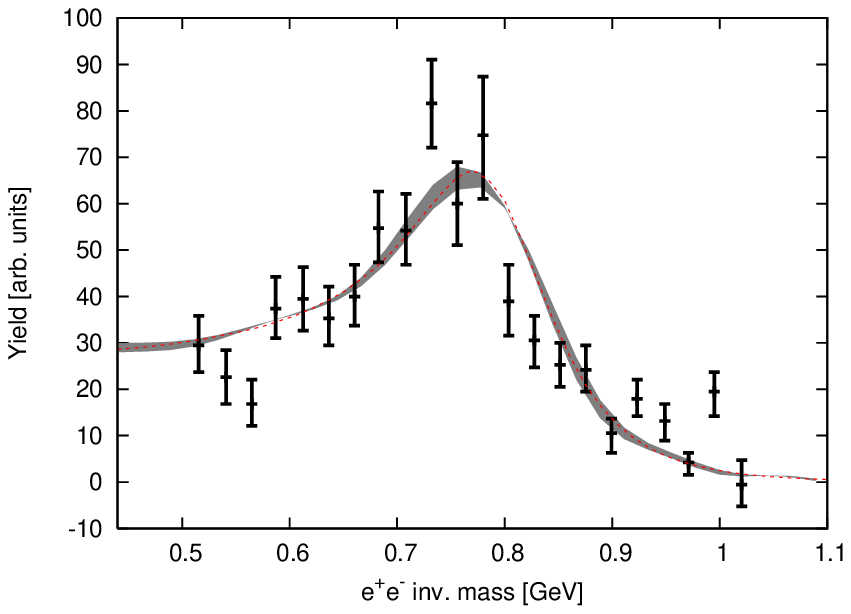}
\end{minipage}
\caption[]{Dilepton spectra measured in nuclear photo-production
by CLAS at JLAB~\cite{Wood:2008ee}, including transport model
calculations~\cite{Muhlich:2002tu} for the decays of  
light vector mesons $\rho$, $\omega$ and $\phi$. Right panel: 
CLAS ``excess spectra" compared to calculations~\cite{Riek:2008ct} 
using the same $\rho$ spectral function as for the NA60 data in
Figs.~\ref{fig_dndm}, \ref{fig_rate}.}
\label{fig_clas}
\end{figure}
The left panel of Fig.~\ref{fig_clas} shows the dilepton signal from 
iron targets, compared to Boltzmann transport 
calculations~\cite{Muhlich:2002tu} for $\rho$, $\omega$ and $\phi$
decays. Since the $\omega$ and $\phi$ peaks are essentially unaffected 
by the medium
(and well concentrated in mass), their contribution can be subtracted  
from the spectrum (much like for the NA60 data) leading to an ``excess'' 
signal as shown by the data in the right panel of Fig.~\ref{fig_clas}. 
Breit-Wigner fits have been applied resulting in a moderate $\rho$
broadening to $\Gamma_\rho^{\rm med}\simeq220$\,MeV~\cite{Wood:2008ee}. 
The $\rho$ spectral function used in the NA60 context has been applied 
to the CLAS experiment in combination with a realistic elementary 
production amplitude and a somewhat schematic modeling of the spatial 
propagation, accounting for the production kinematics and nuclear 
density profile~\cite{Riek:2008ct}. The CLAS data are reasonably well 
described; it turns out that for the Fe target the typical densities 
and 3-momenta probed in the spectral 
function are $\bar{\varrho}\simeq0.5\varrho_0$ and $\bar{q}\simeq2$~GeV. 
The latter are the main reason for moderating the medium effects
in the current CLAS data (recall left panel of Fig.~\ref{fig_cold}).
However, at high momenta additional medium effects not included in the 
spectral function of Ref.~\cite{Rapp:1999us} may occur, see, e.g., 
the discussion in Ref.~\cite{Rapp:1999ej}.

%%%%%%%%%%%%%%%%%%%%%%%%%%%%%%%%%%%%
\section{Open Charm and Transport}
\label{sec_charm}
\begin{figure}[!t]
\begin{minipage}{0.4\linewidth}
\begin{center}
\includegraphics[width=0.65\textwidth]{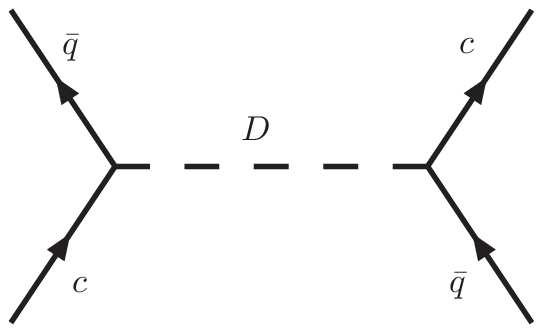}
\includegraphics[width=0.9\textwidth]{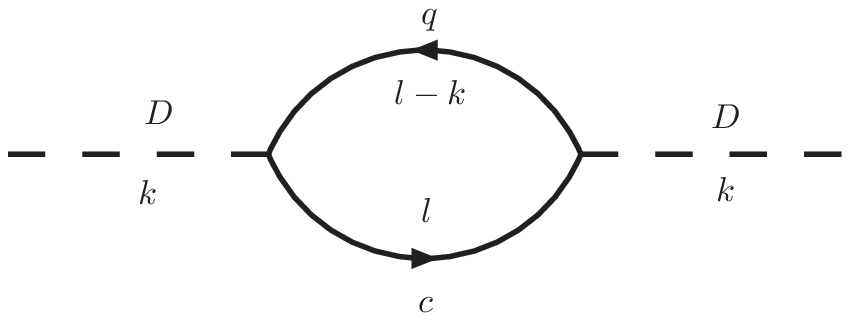}
\end{center}
\end{minipage}
\begin{minipage}{0.6\linewidth}
\includegraphics[width=0.95\textwidth]{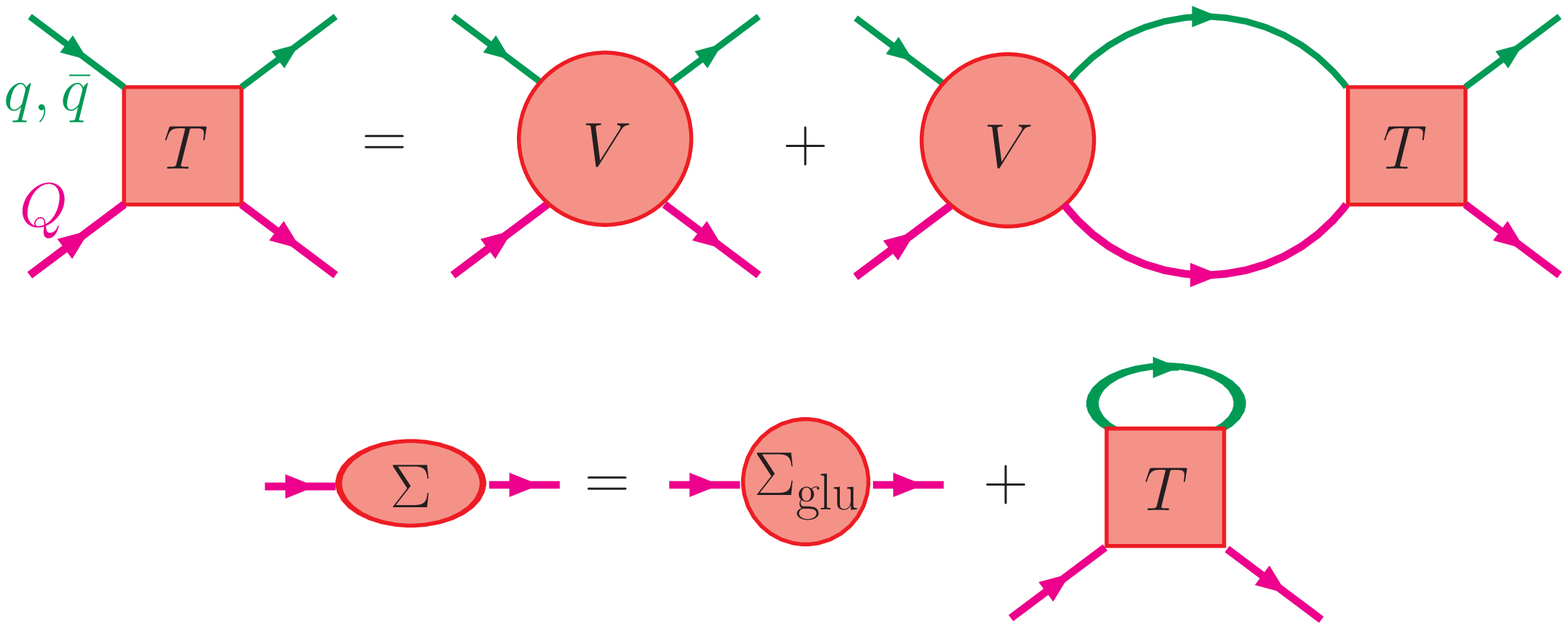}
\end{minipage}
\caption[]{Left panels: $c$-quark scattering off light antiquarks via 
$s$-channel $D$-mesons (upper panel) and pertinent $D$-meson 
selfenergy (lower panel) within the effective resonance model in the 
QGP~\cite{vanHees:2004gq}. Right panels: selfconsistent Brueckner scheme 
for heavy quarks in the QGP based on an in-medium heavy-light $T$-matrix  
with interaction potential $V$ (upper panel) and pertinent HQ
selfenergy (lower panel)~\cite{vanHees:2007me}.}
\label{fig_graph}
\end{figure}

\begin{figure}[!t]
\begin{minipage}{0.5\linewidth}
\vspace{-0.5cm}
\includegraphics[width=0.95\textwidth]{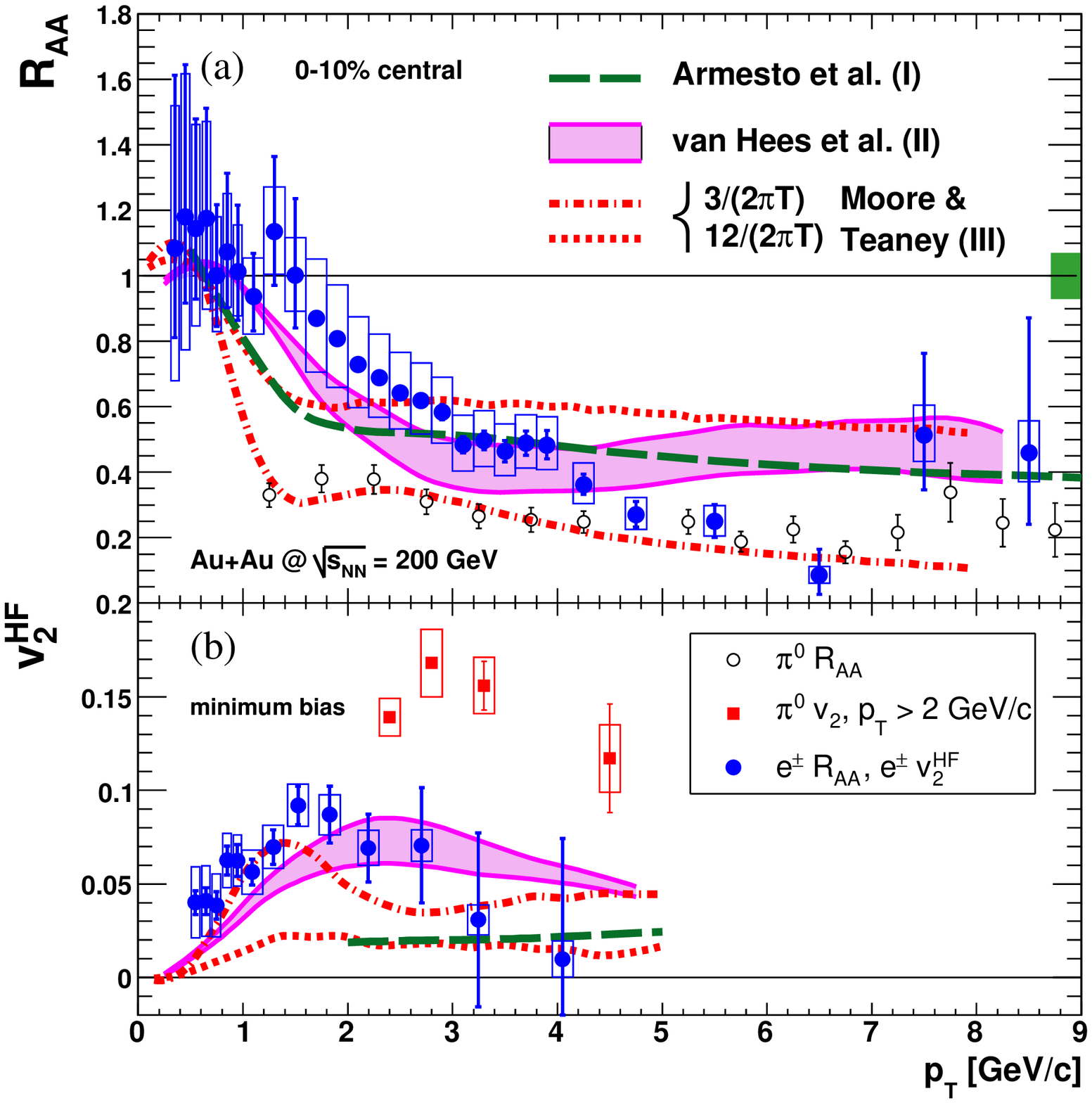}
\end{minipage}
\begin{minipage}{0.5\linewidth}
\includegraphics[width=0.95\textwidth]{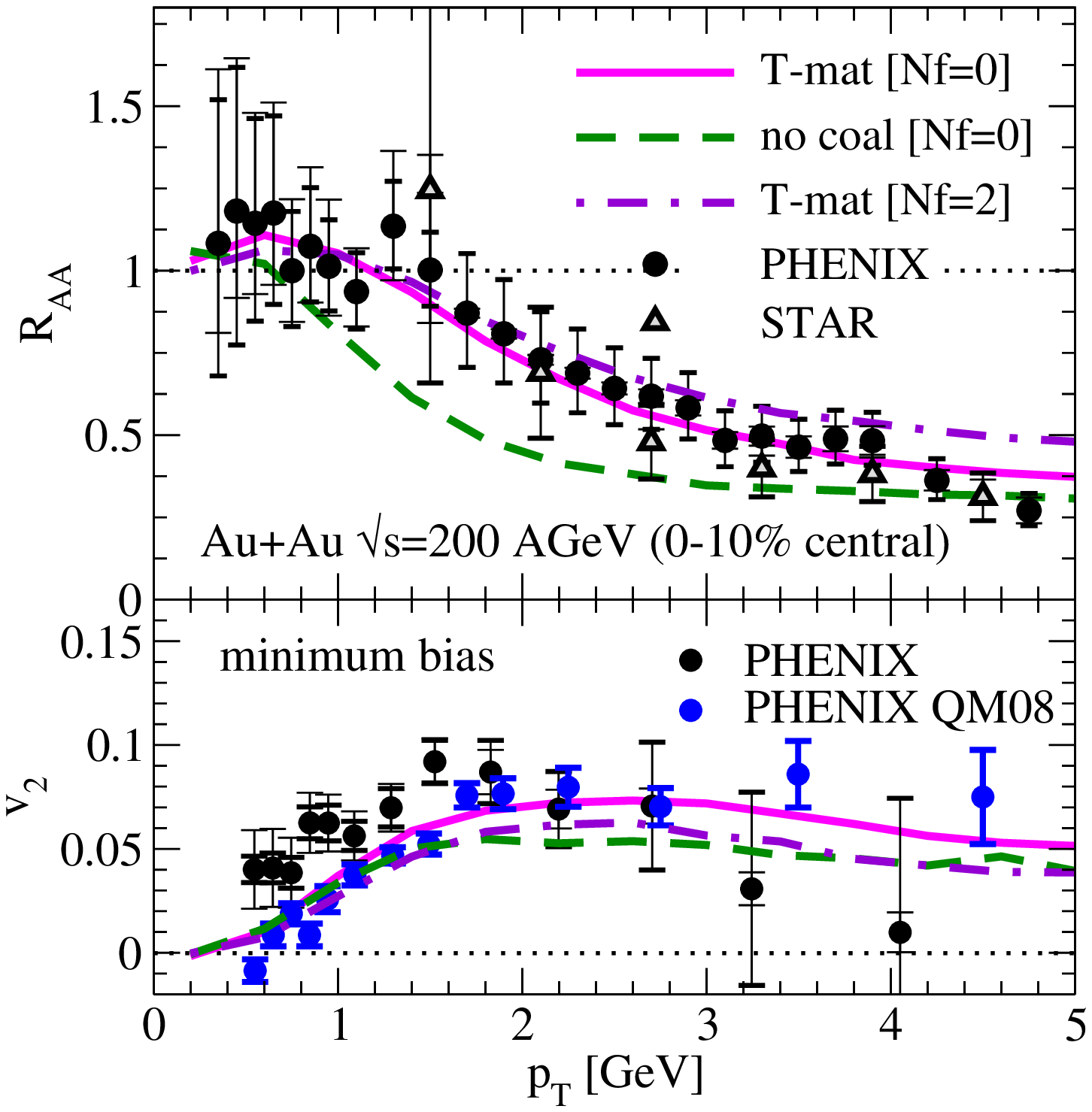}
\end{minipage}
\caption[]{RHIC data~\cite{Adare:2006nq,Abelev:2006db,Hornback:2008ur} 
for the nuclear modification factor (upper panels) and elliptic flow (lower 
panels) of HQ decay electrons in  
Au-Au collisions, compared to theory. 
Left panels: relativistic Langevin simulations using upscaled pQCD 
elastic scattering cross sections (short-dashed and dash-dotted 
line)~\cite{Moore:2004tg} or effective resonances+pQCD elastic 
scattering (bands)~\cite{vanHees:2005wb}, and radiative energy-loss
calculations (long-dashed lines)~\cite{Armesto:2005mz}. 
Right panels: heavy-light quark 
$T$-matrix+pQCD elastic interactions~\cite{vanHees:2007me} using 
internal energies from quenched (solid line) 
and $N_f$=2 (dash-dotted line) thermal lQCD~\cite{Kaczmarek:2003ph}; 
the dashed lines are obtained if hadronization via heavy-light quark 
coalescence at $T_c$ is switched off.  
  }
\label{fig_elec}
\end{figure}

%%%%%%%%%%%%%%%%%%%%%%%%%%%%%%%%%%%%
The masses of charm and bottom quarks (and hadrons) are much larger than 
the typical temperatures realized in heavy-ion collisions at SPS and 
RHIC, $m_{Q}\gg T$ ($Q$=$b$,$c$). Furthermore, a typical momentum transfer
from the heat bath, $|q^2|\simeq T^2$, is parametrically smaller than the 
thermal momentum of $c$ and $b$ quarks, $p^2\sim 3m_{Q}T \gg |q^2|$.  
Thus a diffusion approximation to the Boltzmann equation becomes 
applicable leading to a Fokker-Planck 
equation~\cite{Svetitsky:1987gq,vanHees:2004gq,Moore:2004tg,Mustafa:2004dr}, 
\begin{equation}
\frac{\partial f_Q}{\partial t} = \gamma \frac{\partial (pf_Q)}{\partial p}
+ D \frac{\partial^2 f_Q}{\partial p^2} \ ,  
\end{equation}
for the heavy-quark (HQ) phase-space distribution, $f_Q$. The scattering
rate, $\gamma$, and momentum diffusion coefficient, $D$, are related
via the Einstein relation, $T=D/(\gamma m_Q)$. 
Applications to RHIC data revealed that perturbative QCD (pQCD) elastic 
scattering is insufficient to generate the observed elliptic flow, even 
with a strong coupling constant as large as $\alpha_s=0.4$, supporting 
the notion of a strongly coupled QGP (sQGP). At strong coupling, diagrams 
with large contributions have to be resummed, possibly leading to the 
appearance of collective modes (bound states or resonances). In this spirit, 
the effective resonance model for $c$- and $b$-quark scattering through 
in-medium $D$ and $B$ meson has been introduced~\cite{vanHees:2004gq}. 
The pertinent $s$-channel
diagrams are displayed in the left panels of Fig.~\ref{fig_graph}. An 
approximately 4-fold decrease of the HQ thermalization time, 
$\tau_Q=\gamma^{-1}$, has been found relative to pQCD. 
When implemented into relativistic Langevin simulations within an 
expanding fireball for Au-Au collisions at RHIC~\cite{vanHees:2005wb}, 
the recent data on suppression and elliptic flow of semileptonic decay 
electrons are fairly well described~\cite{Adare:2006nq,Abelev:2006db}, 
cf.~left panel of Fig.~\ref{fig_elec} (see also 
Refs.~\cite{Zhang:2005ni,Gossiaux:2008jv,Akamatsu:2008ge}). 
Heavy-light quark coalescence in the hadronization
process at $T_c$ plays a significant role in increasing {\em both}
$R_{AA}$ and $v_2$.

One may ask whether resonant HQ interactions in the QGP can be
understood more microscopically. This question has been studied  
using a heavy-light quark $T$-matrix equation~\cite{vanHees:2007me},
\begin{equation}
T_{qQ} = V_{qQ} + V_{qQ} \, G_{qQ} \, T_{qQ}
\end{equation} 
diagrammatically depicted in the upper right panel of Fig.~\ref{fig_graph}.
The key input is the driving kernel (potential), $V_{qQ}$, which
has been assumed to coincide with the static heavy-quark internal 
energy computed in thermal lQCD, augmented by relativistic corrections
due to color-magnetic current-current interactions. In addition
to the color-singlet channel, the $Q$-$\bar{q}$ and $Q$-$q$ interactions
in the color-octet, -antitriplet and -sextet channels have been estimated
(using Casimir scaling). It turns out that the interaction
strength is concentrated in the attractive singlet and antitriplet
channels, supporting Feshbach-like meson and diquark resonances close to 
the $q$-$Q$ threshold up to temperatures of $\sim$1.7\,$T_c$ and 
$\sim$1.4\,$T_c$, respectively. Compared to the resonance model, the HQ 
interaction in the $T$-matrix approach of similar strength close 
to $T_c$, but weakens at higher $T$ due to color-screening in the
potential which dissolves the resonances.
An open problem remains the nonperturbative treatment of HQ-gluon 
interactions. The application of the $T$-matrix approach to RHIC 
electron data looks promising (right panels in Fig.~\ref{fig_elec}), 
but significant uncertainties remain which currently inhibit definite 
conclusions about the microscopic origin of HQ diffusion. 

What effects can be expected for charm-quark diffusion at finite 
$\mu_q$, relevant for CBM energies and a RHIC energy scan? 
The effective resonance model suggests that, in a quark-dominated
environment, anticharm quarks ($\bar{c}+q\to \bar{D}$) interact more
frequently than charm quarks ($c+\bar{q}\to D$). However, in the
$T$-matrix approach, scattering via (anti-) diquarks, $cq$ and 
$\bar{c}\bar{q}$, is equally important, thus washing out the 
asymmetry. Moreover, in the hadronic phase, the baryon excess 
favors $D$-meson scattering via $\Lambda_c N^{-1}$ excitations 
over its antiparticle conjugate.
A promising possibility could be the development of the $\sigma$ soft 
mode close to the critical point, which is particularly pronounced
in the spacelike regime~\cite{Fujii:2003bz}. {\em If} charm quarks
couple to the $\sigma$, their $t$-channel exchange cross section with
light quarks (and hadrons) could be ``critically" enhanced leaving 
observable traces in $D$-meson $p_T$ spectra and $v_2$ (see 
Ref.~\cite{Zhuang:1995uf} for a related study in the light-quark 
sector). 

%%%%%%%%%%%%%%%%%%%%%%%%%%%%%%%%%%%%%%%%%%%%%%%%%
\section{Charmonia: Screening and Dissociation}
\label{sec_charmonium}
%%%%%%%%%%%%%%%%%%%%%%%%%%%%%%%%%%%%%%%%%%%%%%%%%
The dissolution temperature of charmonia in medium largely depends 
on two mechanisms: color-screening of the inter-quark force and 
inelastic dissociation reactions. While the 
former largely governs the $Q$-$\bar{Q}$ binding energy, $\varepsilon_B$ 
(via spacelike gluon exchange), the latter determines the inelastic 
width, $\Gamma_\psi^{\rm inel}$, of the bound-state (via dissociation 
reactions with timelike (on-shell) partons in the heat bath). 
Within a schematic pole ansatz, both mechanisms figure into the 
charmonium spectral function as
\begin{equation}
\sigma_\psi(\omega) \sim {\rm Im}~D_\psi(\omega) \sim
{\rm Im}~[\omega-2m_c^* +\varepsilon_B + i\,\Gamma_\psi/2 ]^{-1} 
\label{sig-psi}
\end{equation}
(a pole ansatz applies to a well-defined bound-state/resonance).
Note the dependence on the in-medium $c$-quark mass, $m_c^*$, and 
that the total width, $\Gamma_\psi$, includes contributions from elastic 
scattering as well. However, as we will see below, binding energies and 
dissociation reactions mutually influence each other.

Recent years have seen a revival of potential models to evaluate 
in-medium quarkonium 
properties~\cite{Mocsy:2007yj,Cabrera:2006wh,Alberico:2006vw,Wong:2006bx,Laine:2007qy,Brambilla:2008cx}. 
\begin{figure}[!t]
\begin{minipage}{0.5\linewidth}
\includegraphics[width=1.0\textwidth]{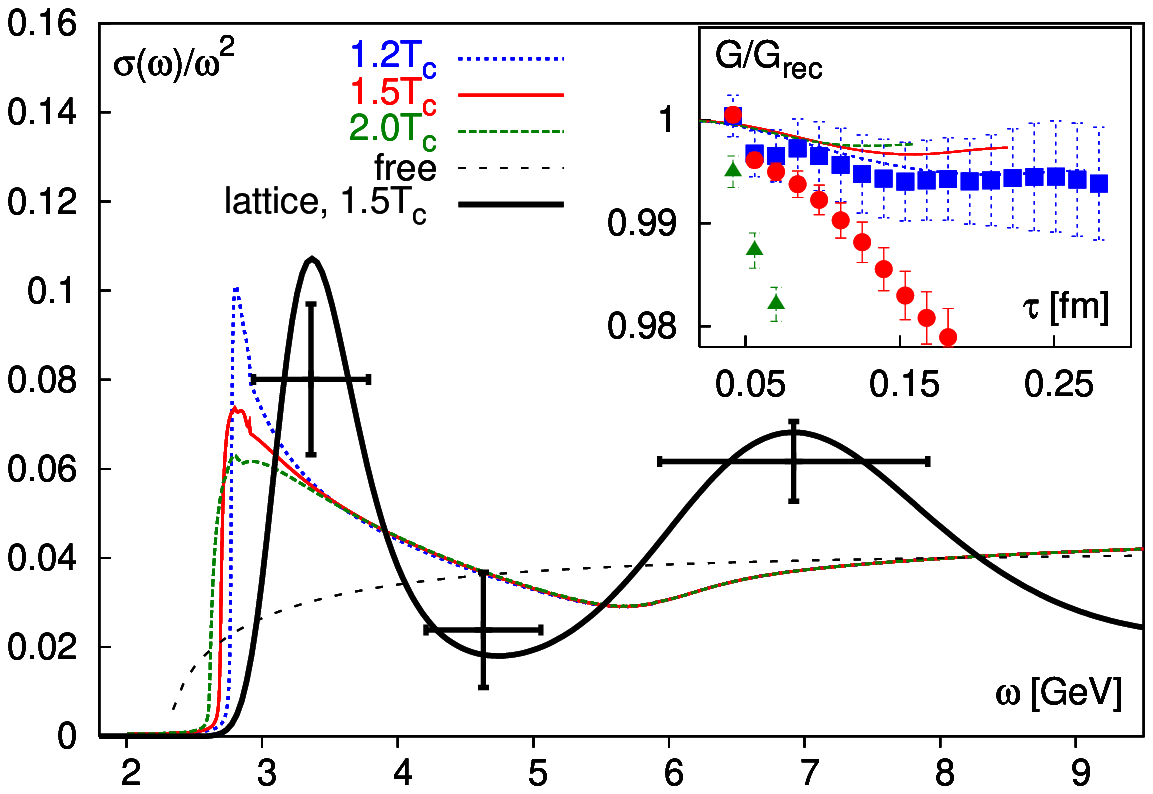}
\end{minipage}
\begin{minipage}{0.5\linewidth}
\includegraphics[width=0.9\textwidth]{ImG-etac-CR.eps}
\includegraphics[width=0.9\textwidth]{RG-etac-CR.eps}
\end{minipage}
\caption[]{$S$-wave $c$-$\bar{c}$ spectral functions and Euclidean-time
correlators (normalized with a ``reconstructed" correlator, $R_G(\tau)\equiv
G(\tau)/G_{\rm rec}(\tau)$) using a HQ potential close to the lQCD free
energy (left panels)~\cite{Mocsy:2007yj} or corresponding to the
lQCD internal energy (right panels)~\cite{Cabrera:2006wh}.}
\label{fig_corr}
\end{figure}
This was largely spurred by the hope that the input potential can be
taken in a model-independent way from the HQ free energy computed in
thermal lQCD, and that resulting spectral functions can be discriminated
via comparisons to Euclidean correlation functions independently
computed in lQCD. There is, however, an ongoing controversy as to
which quantity to identify with a potential (free vs. internal energy),
and concerning the gauge variance of the projection on the color-singlet
channel~\cite{Philipsen:2008qx}. More quantitatively, the current 
situation is illustrated in Fig.~\ref{fig_corr}.
Roughly speaking, when the singlet free energy, $F_{Q\bar{Q}}^1(r,T)$, 
is used as potential, ground-state charmonia dissolve at temperatures 
below 1.5\,$T_c$ (left panel)~\cite{Mocsy:2007yj}. On the other hand, 
with the internal energy, 
$U_{Q\bar{Q}}^1 =F_{Q\bar{Q}}^1(r,T) +T S_{Q\bar{Q}}^1(r,T)$ 
($S_{Q\bar{Q}}$: entropy), a $J/\psi$ peak in the 
spectral function can survive up to 2.5-3\,$T_c$ (upper right 
panel)~\cite{Cabrera:2006wh}.  
The spectral functions have been used to calculate temporal Euclidean 
correlation functions, 
\begin{equation}
G_\psi(\tau,p;T)=
\int\limits_0^\infty {d\omega} \ \sigma_\psi(\omega,p;T) \
\frac{\cosh[(\omega(\tau-1/2T)]}{\sinh[\omega/2T]}  \ . 
\label{G-tau}
\end{equation}
which are usually normalized to a ``reconstructed" correlator, 
$G_{\rm rec}$, computed with a vacuum spectral function (but
identical finite-temperature kernel in eq.~(\ref{G-tau})).
Surprisingly, both weak and strong-binding scenarios for the 
spectral function result in correlator ratios which are around 
one and depend weakly on temperature (compatible with lQCD 
results~\cite{Datta:2003ww,Jakovac:2006sf,Aarts:2007pk}), cf.~inset 
in the left panel and lower right panel of Fig.~\ref{fig_corr}.
The reason for this ``redundancy" is the underlying effective quark
mass, which is calculated from the asymptotic value of the HQ
free or internal energy, $m_c^*=m_c^0+\Delta m_c$ with 
$2\Delta m_c = F(r\to\infty,T)$ or $U(r\to\infty,T)$. 
For the free energy, $\Delta m_c$ is substantially reduced from its
vacuum value, thus lowering the in-medium $c\bar{c}$ threshold
considerably; this compensates for the lack of a bound state in
providing low-energy strength in the spectral function to ensure
a stable correlator ratio. For the internal energy, $\Delta m_c$
is significantly larger which, together with a stronger binding, 
leads to an essentially stable $J/\psi$ peak position and thus a 
roughly stable $T$-dependence of the correlator ratio. 
More work is needed to disentangle these two scenarios. 
Finite-width effects on the correlators have received rather little 
attention thus far, but they seem to further stabilize 
the $T$-dependence~\cite{Cabrera:2006wh}.
If a reliable understanding of quarkonium correlators in the QGP 
at $\mu_q=0$ in terms of potential models can been established, 
it might serve as a benchmark to extrapolate, e.g., color-screening
effects to finite $\mu_q$.  

\begin{figure}[!t]
\begin{minipage}{0.5\linewidth}
\includegraphics[width=0.95\textwidth]{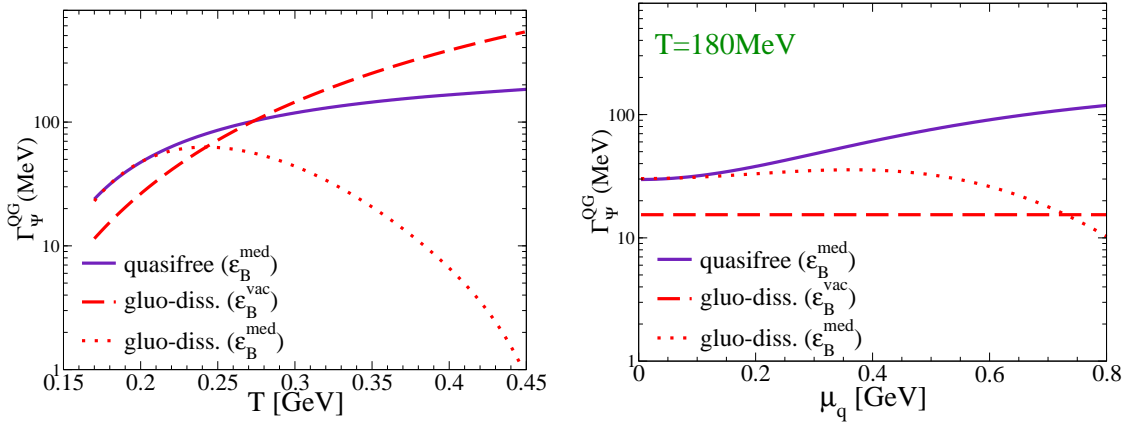}
\end{minipage}
\begin{minipage}{0.5\linewidth}
\includegraphics[width=0.95\textwidth]{gam-psi-muq.eps}
\end{minipage}
\caption[]{$J/\psi$ dissociation rate (=inelastic width) in a QGP as a 
function of $T$ (at $\mu_q$=0, left panel) and $\mu_q$ (at $T$=180~MeV, 
right panel) for gluo-dissociation ($g+J/\psi\to c+\bar{c}$) and 
quasifree destruction ($p+J/\psi\to p+c+\bar{c}$ with $p=q,\bar{q},g$)
with vacuum and in-medium reduced $J/\psi$ binding energy.}
\label{fig_gam-psi}
\end{figure}
The dissociation width of heavy quarkonia in medium is pivotal 
for a quantitative description of suppression (and
regeneration!) reactions in HICs. The prevalent dissociation 
mechanism in the QGP depends on the binding energy of the bound 
state~\cite{Kharzeev:1995ju,Grandchamp:2001pf,Park:2007zza,Zhao:2007hh}, 
cf.~left panel of Fig.~\ref{fig_gam-psi}.
For large binding, $\varepsilon_B\sim 3T$, thermal-gluon absorption is
most efficient, $g+J/\psi\to c+\bar{c}$ (and formally the process to
leading-order in $\alpha_s$). For small binding, $\varepsilon_B < T$,
the phase space for gluo-dissociation shrinks leading to a decrease
in its rate with $T$. Thus, ``quasi-free'' dissociation, 
$p+J/\psi\to p+c+\bar{c}$, albeit formally of higher order in 
$\alpha_s$, takes over. Note that the quasi-free process can be
induced by both gluons and anti-/quarks. This has consequences at 
finite $\mu_q$, in that quasifree dissociation is additionally enhanced 
over gluo-dissociation due to an increasing abundance of thermal 
quarks, cf.~right panel of Fig.~\ref{fig_gam-psi}~\cite{Zhao:2009}.

%%%%%%%%%%%%%%%%%%%%%%%%%%%%%%%%%%%%
\section{Conclusions}
\label{sec_concl}
%%%%%%%%%%%%%%%%%%%%%%%%%%%%%%%%%%%%
Instead of a formal summary of this paper, let us reiterate what we 
find the most promising perspectives regarding the finite-$\mu_q$ 
dependence of dileptons and charm/onia ath this point.  
For low-mass dileptons, we have identified the mass region around
$M\simeq0.2$~GeV as the most sensitive one for baryon-driven medium 
effects. In addition, with a good knowledge of the in-medium spectral 
shape of the EM correlator, the dilepton yield can finally be used for 
an accurate determination of the fireball lifetime in HICs, which might
be useful in detecting (the onset of) an extended quark-hadron mixed 
phase. For open charm, a ``critical" enhancement of $\sigma$ exchange
in $c$-quark or $D$-meson scattering in the vicinity of the critical 
point may occur, potentially affecting transport properties in an 
observable way ($p_t$ spectra and elliptic flow). For charmonia, a 
hope is to establish the validity of finite-$T$ potential models and 
extrapolate them to finite $\mu_q$, augmented by microscopic 
calculations of dissociation rates.  
These developments are particularly exciting in view of future tests
in several heavy-ion programs around the world.

\section*{Acknowledgments} 
It is a pleasure to thank D.~Cabrera, V.~Greco, M.~Mannarelli, F.~Riek, 
H.~van Hees, J.~Wambach and X.~Zhao for their collaboration on various 
aspects of the presented topics, and H.~van Hees for a careful reading 
of the ms.
This work has been supported by a U.S. National Science Foundation 
CAREER award under grant no. PHY-0449489 and by the A.-v.-Humboldt 
Foundation (Germany) through a Bessel award.

\end{document}